\documentclass{style/nseJournal}

\usepackage{graphicx} 
\usepackage{hyperref}
\usepackage{multirow}
\usepackage{array}
\usepackage{titlesec}
\usepackage{fancyhdr}
\usepackage[export]{adjustbox}
\usepackage{siunitx}
\usepackage{tabularx}
\usepackage{subcaption}
\usepackage[numbers]{natbib}
\usepackage{wrapfig}

\begin{document}

\title{Heavy Water Displacement in Molecular Sieve Drying Beds at Various Humidities} %title of paper

% Use the \addAuthor macro to add authors in the order they should appear. The second argument corresponds to
% the affiliation declared below.
% The corresponding author should be wrapped in \correspondingAuthor
\addAuthor{\correspondingAuthor{B. Massett}}{a}
\correspondingEmail{bmm8532@rit.edu}
% The corresponding author's email can be specified using \correspondingEmail
\addAuthor{W. T. Shmayda}{b}

% Affiliations can be added in the order they should appear. For breaks in addresses, use either \\ or \tabularnewline
\addAffiliation{a}{Rochester Institute of Technology}
\addAffiliation{b}{Tritium Solutions Inc.}

% Add keywords to appear in Abstract in the order they should appear
\addKeyword{Dryer Beds}
\addKeyword{Heavy Water}
\addKeyword{Tritiated Water}
\addKeyword{Displacement}

\titlePage

\begin{abstract}
    Tritium plays a critical role in nuclear fusion power plant designs and dryer beds are an essential tool for managing tritiated water vapor. A series of tests were performed to investigate the ability of a saturated dryer to preferentially adsorb heavy water vapor. The design of passive tritiated control systems is feasible by utilizing a dryer's ability to preferentially trap heavier isotopologues of water. This work investigates this displacement phenomenon and the effect of the heavy water humidity on the dryers performance. Significant displacement was observed when a humid stream of heavy water was diverted through a dryer pre-saturated with light water, as indicated by changes in the partial pressures of $D_2O$ and $H_2O$. After the capture of heavy water in the bed, the subsequent rise in $D_2O$ partial pressure depended on the humidity of heavy water in the gas stream. Higher humidity values lead to faster and steeper mass transfer profiles in the dryer, which could be empirically fit with sigmoid curves.
\end{abstract}

\section{Introduction}
    Tritium is an important nuclide in the deployment of potential nuclear fusion power plants. The tritium-deuterium cross section for fusion reactions is the largest of all the fusion reaction options envisioned. Additionally, these high reaction cross-sections occur at the lowest plasma temperatures, providing favorable fusion prospects. Alongside this, tritium can be produced via neutron bombardment of lithium and/or lithium compounds. High-energy neutrons from the fusion reaction interact with the breeding medium to produce more tritium, which in turn provides more tritium for the D-T reaction. Tritium provides the glue that couples the nuclear reaction to the consumption of lithium in the deuterium-lithium fuel cycle.

    With tritium acting as both a reactant and a product, a closed-loop of tritium burning and production is possible within a fusion machine. This makes proper handling of tritium very important. If tritium leaks into the environment, it could cause concerns over health problems \citep{Matsumoto}. 

    \begin{wrapfigure}{l}{0.4\textwidth}
        \centering
        \includegraphics[width=0.4\columnwidth]{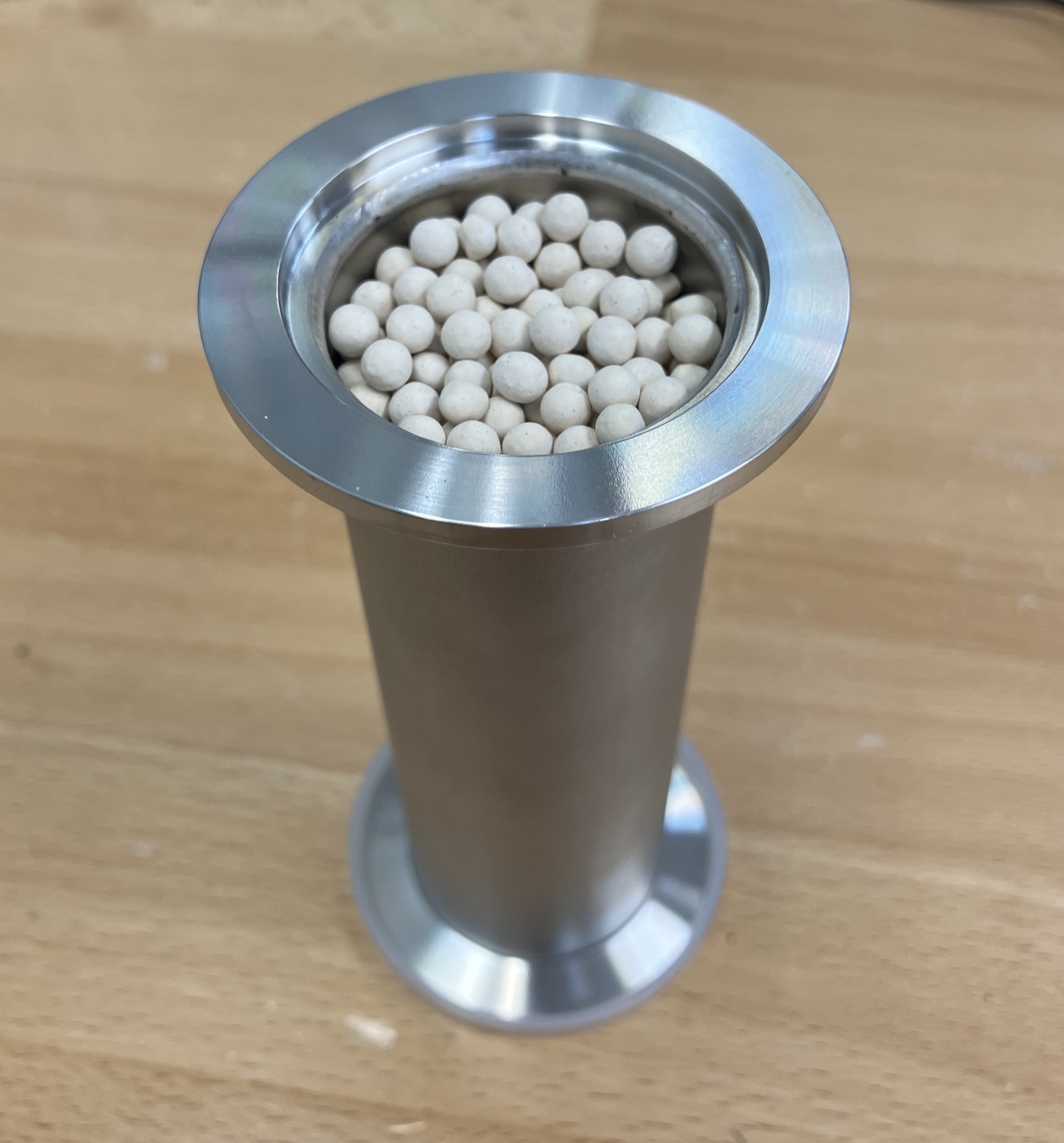}
        \caption{Example of packed dryer bed used in this study}
        \label{fig:pack_bed_example}
    \end{wrapfigure}

    Public acceptance of fusion machine will be driven in part by demonstrating that tritium emissions are mitigated effectively. To avoid contaminating the environment surrounding a fusion machine, the majority of the tritiated effluents need to be captured with the intent of recovering the tritium. A robust method is to convert all effluent to tritiated water and to capture the water on adsorbent materials. Studies on different adsorbents properties and efficiencies are critical to properly design air-detritiation systems.

    Experimental results with protium and deuterium are important stepping stones in the development of air-detritiation systems. Insights on the behavior of tritium can be deduced from the behavior of deuterium relative to protium in experimental air-detritiation systems. These results act to inform future tritium-based experiments and air-detritiation designs.

    A multitude of porous materials have been found and studied for extracting water vapor from carrier streams. These materials are typically contained between two filters in a vessel to form a `fixed-bed' configuration through which the carrier can flow. The most common adsorbents used are silica gel, activated alumina, and molecular sieves. An example of the fixed bed we used in this study can be seen in Fig. \ref{fig:pack_bed_example}.
    
    The adsorbent material selected for this study is $4A$ molecular sieve. This material is an extremely uniform and porous zeolite. This gives the molecular sieve the ability to adsorb water vapor on its surface. For molecular sieve 4A, maximum loading of water vapor can be around 22wt\% \citep{UOPpamphlet}. The drying efficiency of these packed-beds can reach dew points of $-95$°$C$ under ideal conditions, which correlates to 35ppb at 1atm. As well, the water can easily be extracted from the molecular sieve through heating \citep{DryersSpecs}. The efficiency and usefulness of these adsorption beds have also been studied and confirmed with tritiated water \citep{NAKASHIMA,Tanaka}.

    The important property of these drying beds is the ability for heavier water species to displace $H_2O$ already adsorbed in a saturated bed. This allows fully saturated dryer beds to passively stay in a gas system while preserving the ability to capture tritiated water. This has been observed and characterized with multiple adsorbent materials \citep{osti_4734021}. This can also be observed with deuterated water, which can be extrapolated to tritiated water as the adsorption characteristics have been shown to be quite similar \citep{Nishikawa}. Tritium applications would involve very low concentrations, which yields $HTO$ as the dominant active contaminant. Using $D_2O$ is the ideal analog to $HTO$ since both water species have a molecular mass of 20 amu. The properties of this displacement can be characterized with varied heavy water concentration flowing into the bed, similar to MTZ tests with $H_2O$.

\section{Theory}
\label{sec:theory}

    Understanding the properties of a drying bed in order to scale it to industrial applications requires an understanding of the mass transfer zone (MTZ) and how it changes with superficial velocity. The superficial velocity ($V$) is defined as

    \begin{equation}
        V = \frac{F_{rate}}{A}
        \label{eq:Velocity}
    \end{equation}

    where $F_{rate}$ is the volumetric flow rate, and $A$ is the cross sectional area of the dryer bed. Every trial carried out in the tests discussed in this report used the same bed with a cross-sectional area of $22.86\: cm^{2}$.

\begin{wrapfigure}{r}{0.55\textwidth}
        \centering
        \includegraphics[width=0.55\columnwidth]{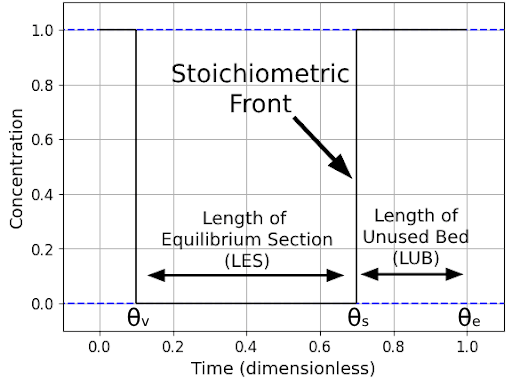}
        \caption{Idealized stoichiometric front}
        \label{fig:stoich}
    \end{wrapfigure}

    Under ideal, and initially dry, fixed-bed conditions, adsorption is defined by three separate limiting cases. These are: negligible external/internal transport resistances, ideal plug flow, and an adsorption isotherm beginning at the origin \citep{Seader}. These idealizations lead to an instantaneous change in the concentration of the effluent from near zero to equal that of the feed once the bed is saturated. This front is referred to as the stoichiometric front. As this front moves through the bed, it divides the bed into two sections. The section upstream of the front is the saturated portion of the bed, which is in equilibrium with the feed flow ($LES$). Downstream of the front represents the unused portion of the bed ($LUB$). The stoichiometric front, $LES$, and $LUB$ are labeled on a simple, idealized curve in Fig. \ref{fig:stoich}. 

    $\theta_{v}$, $\theta_{s}$, $\theta_{e}$, represent the time equal to the time the bed is valved in, the time the stoichiometric front exits the bed, and the time at the end of the experiment, respectively. In the ideal case the $LUB$ is equal to zero because the stoichiometric front is infinitely thin. 

    This idealized example represents the dew point reading downstream of a dryer bed. The bed is valved into the loop at $\theta_{v}$, sharply reducing the effluent humidity. The effluent dew point stays low until the adsorption capacity of the bed is reached. At this time, $\theta_{s}$, the effluent humidity sharply increases back to the inlet concentration value. This sharp increase is referred to as the stoichiometric front.

\begin{figure}[!h]
        \centering
        \includegraphics[width=0.8\columnwidth]{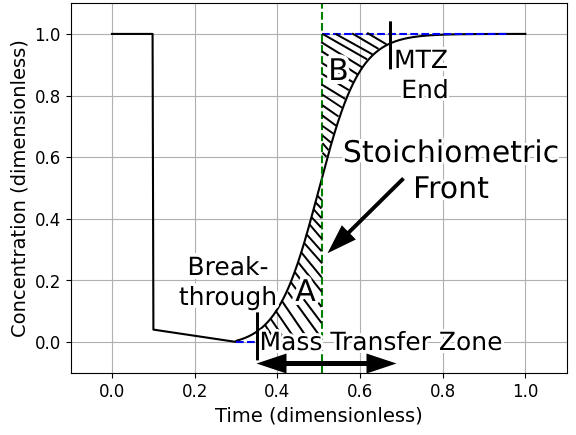}
        \caption{Stoichiometric of Realistic Breakthrough Curve}
        \label{fig:RealStoich}
    \end{figure}

    A graphical example of the stoichiometric front for non-ideal adsorption is shown in Fig. \ref{fig:RealStoich}. The stoichiometric front for a more realistic profile is located in the MTZ so that the area between the MTZ profile and the front, area $A$, must be equal to area between the front and the MTZ on the downstream side, area $B$, in the figure. In this case the $LUB$ is greater than zero.

    In breakthrough curve data collected experimentally, a gradual curve develops from the minimum to the maximum concentration as the bed is becoming saturated. This realistic situation can be represented by the idealized stoichiometric front. Total water adsorbed can be easily calculated using this equivalent stoichiometric front.
    
    By definition, the stoichiometric front would be at the mid point of a symmetric MTZ. For unsymmetrical MTZs, the stoichiometric front is located at the dimensionless time when the area under the breakthrough curve upstream of the front equals the area above the breakthrough curve downstream of the front.
    
    One property of the adsorption data of interest is the length of the MTZ. This length is calculated from the time needed for the dew point profile to evolve from $2\%$ above background to the $98\%$ of the final dew point value as illustrated in Fig. \ref{fig:RealStoich}. It can be simply calculated as:

    \begin{equation}
        L_{MTZ} = \frac{(\theta_{e}-\theta_{b})}{(\theta_{e}-\theta_{v})}L_{0}
        \label{eq:L_MTZ} 
    \end{equation}

    \noindent where $\theta_{e}$ is the time the MTZ reaches $98\%$, $\theta_{b}$ is the time of MTZ breakthrough ($2\%$), $\theta_{v}$ is the time the bed is valved in, and $L_{0}$ is the length of the whole bed. The numerator is the time duration of the MTZ, and the denominator is simply the time duration from valving in of the bed to the end of the MTZ.

    Khol \citep{Seader} proposed the length of the MTZ depends on the superficial gas velocity ($V$) and a factor ($Z$) which depends on the size of the molecular sieve medium:

    \begin{equation}
        L_{MTZ} = (4975 V)^{0.3}(Z)
        \label{eq:L_MTZ_Theory}
    \end{equation}

    \noindent where $V$ is the superficial velocity in $cm/sec$ and the length of the MTZ is in $cm$. 

    When displacing light water in a bed with heavier isotopologues, the heavy water will eventually gradually increase back to the inlet value, akin to the mass transfer zone concept for initially loading a bed. As such, heavy water displacement can be similarly experimentally characterized by focusing on the downstream humidity response.

    Sigmoid curves provide a rigorous method of characterizing changes in the profiles of the MTZ for different experimental conditions. These curves use two parameters, $k$ and $x_{0}$, within the equation provided below. The value of $x_0$ shifts the sigmoid along the abscissa, showing how the stoichiometric front position changes with superficial velocity, while the value of $k$ indicates the steepness of the sigmoid curve. The MTZ profile needs to be dimensionless to evolve from zero to 1 over time before fitting can take place. With these considerations, we can fit our breakthrough curves as:
    
    \begin{equation}
        S(x) = \frac{1}{1+e^{-k\left(x-x_0\right)}}
        \label{eq:Sigmoid}
    \end{equation}

\section{Experimental Setup}
\begin{figure}[ht]
    \centering
        \includegraphics[width=\columnwidth]{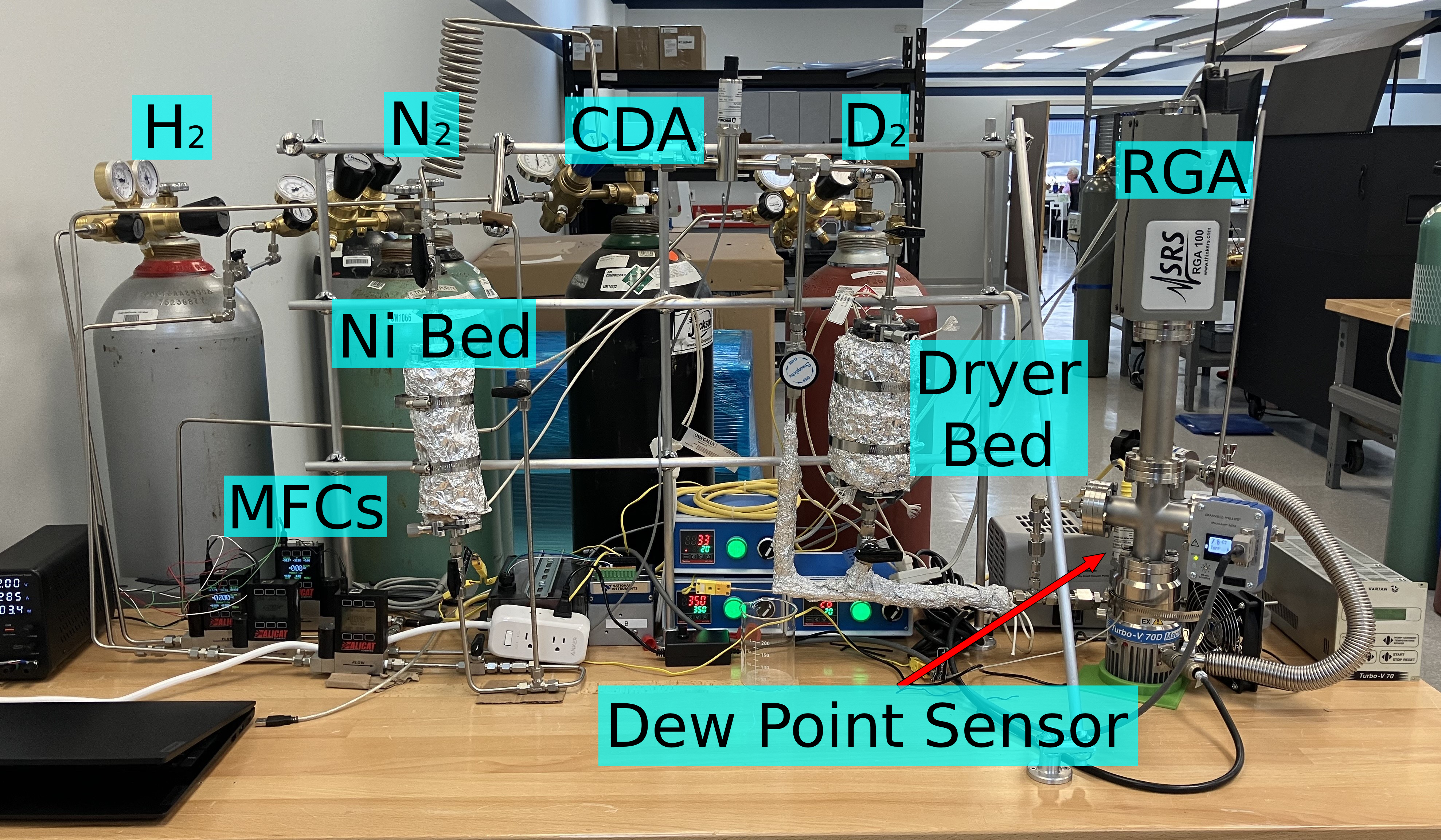}
        \caption{Photograph of the Experimental Setup}
        \label{fig:setup}
    \end{figure}

    An experimental setup was constructed, as shown in Fig. \ref{fig:setup}. It consisted of four separate input lines, each equipped with a mass flow controller, supplying different gases: hydrogen, deuterium, nitrogen, and compressed dry air. The first dew point sensor was installed upstream of the drying bed. The dryer bed contained $155\: g$ of 4A molecular sieve with an internal diameter of $1.43$ inches and a length of $8.0$ inches between the two stainless steel filters. The bed was mounted vertically, with a bypass line around it. The second dew point sensor was positioned downstream of the bed, just before the loop exhaust.

\begin{figure}[ht]
        \centering        
        \includegraphics[width=1\columnwidth]{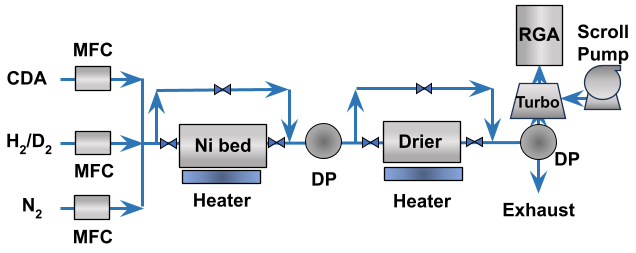}
        \caption{The dryer loop comprises three Mass Flow Controllers (MFC), a heated nickel bed, two Dew Point Sensors (DP), a dryer, vacuum pumps, RGA, and Compressed Dry Air (CDA), hydrogen, deuterium, and nitrogen gas supplies}
        \label{fig:diagram}
    \end{figure}

    The molecular sieve drying bed was regenerated before each experimental trial to restore its adsorption capacity. This was achieved by heating the bed to $350\:^{o}C$  and flowing nitrogen gas through it at $100\:sccm$ overnight. During this process, the second dew point sensor was removed from the setup to prevent potential damage by the hot, humid gas. Fig. \ref{fig:diagram} provides a cartoon of the experimental setup.

   This setup used a nickel bed to produce and control the humidity of the nitrogen carrier during experimental trials. The flow rates of the hydrogen and compressed air through the heated nickel bed were adjusted to produce a stable, pre-determined humidity in the carrier. This setup enabled us to study the effects of varied carrier dew points at fixed flow rates.

   A heat dissipation coil was added above the nickel bed to prevent overheating the upstream dew point sensor. At high flow rates, a substantial quantity of heat was transferred from the hot nickel bed to the carrier stream which in turn warmed the dew point sensor and skewed the DP reading.

   A vacuum system with a residual gas analyzer (RGA) was added to the apparatus to distinguish between the water isotopologues $H_2O$ (18 amu), $HDO$ (19 amu), and $D_2O$ (20 amu). The vacuum system was connected to the downstream line via a sampling pipe equipped with a 5-micron pinhole gasket to restrict flow.

\section{Results and Discussions}
\begin{wrapfigure}{l}{0.45\textwidth}
        \centering
        \includegraphics[width=0.45\columnwidth]{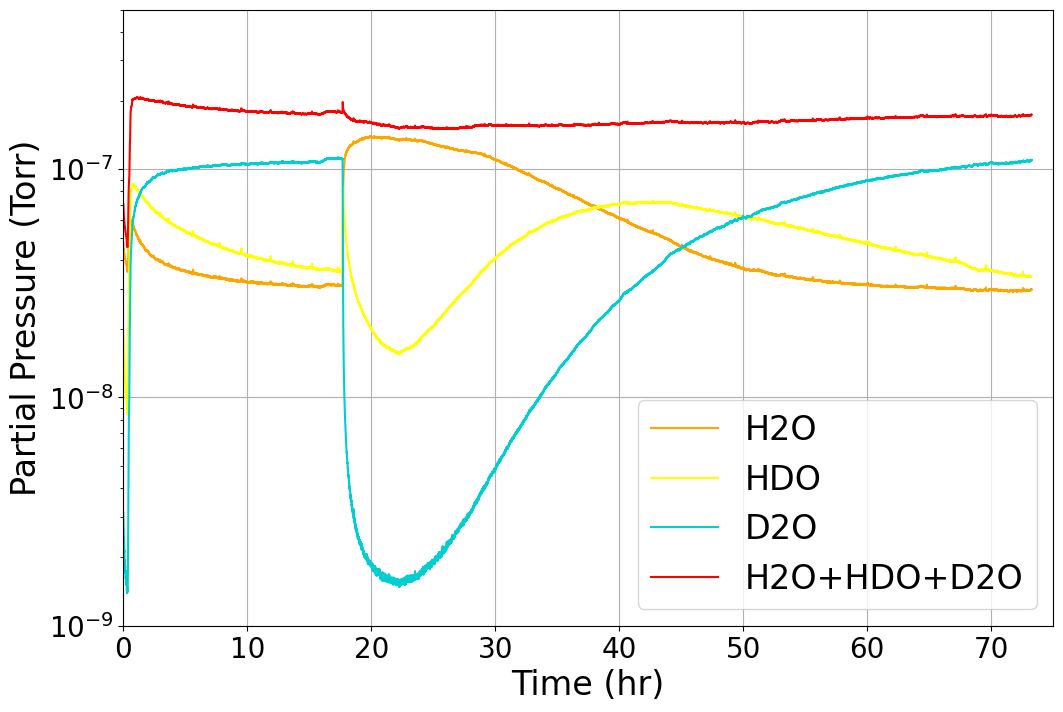}
        \caption{$D_2O$ Displacement Trial 1 at 3.8°C $D_2O$ Dew Point}
        \label{fig:D2O_Trial1}
    \end{wrapfigure}

    Multiple tests were conducted by flowing a nitrogen stream containing heavy water vapor through a dryer bed pre-saturated with light water. Every experiment used a superficial velocity of $5.2\: cm/sec$ while the concentration of $D_2O$ was varied. $D_2O$ dew points varied from $-5\:^{o}C$ up to $10\:^{o}C$ to investigate the effect of heavy water concentration on the displacement behavior. An example of RGA partial pressure data from an experimental trial is shown in Fig. \ref{fig:D2O_Trial1}. The summation of the three tracked water isotopologues is also plotted, revealing a nearly constant curve, indicating a one-to-one displacement phenomenon.

    To accurately determine the partial pressure of $D_2O$ in the gas stream, residual gas analyzer (RGA) readings were allowed to stabilize over a 24-hour period. During this time, the flow rates were held constant and the dryer bed was bypassed. This duration was sufficient to achieve steady-state conditions. Once stability was confirmed, displacement began by valving in the dryer bed saturated with light water.

    The RGA was set to scan at three different molecular masses that correspond to water species: 18 amu ($H_2O$), 19 amu ($HDO$), and 20 amu ($D_2O$). With displacement in the bed being a one-to-one event, the sum total of these water partial pressures should stay relatively constant throughout the duration of the experiments, as observed in Fig. \ref{fig:D2O_Trial1}.

    \begin{wrapfigure}{l}{0.55\textwidth}
        \centering
        \includegraphics[width=0.55\columnwidth]{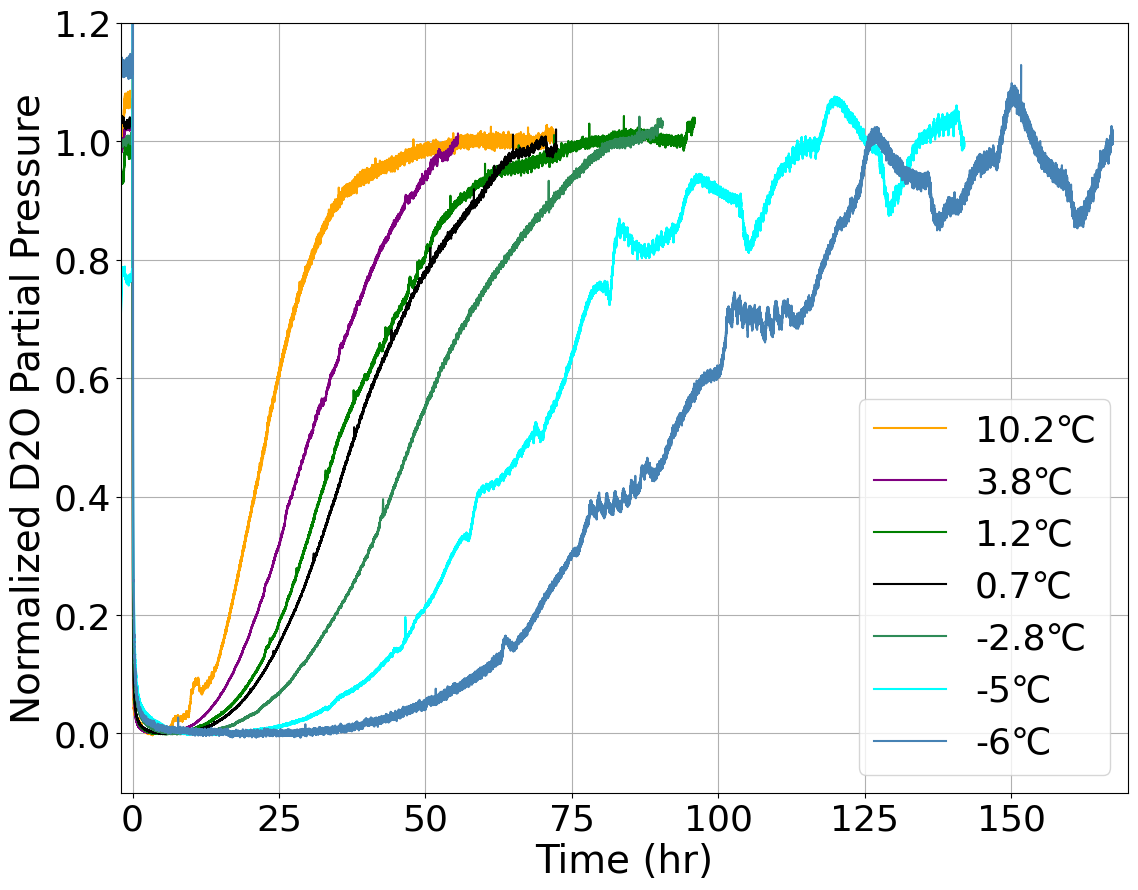}
        \caption{Normalized $D_2O$ Partial Pressure Data from all Displacement Runs Plotted Together}
        \label{fig:All_D2O}
    \end{wrapfigure}

    The evolution of the $D_2O$ profiles for all the runs are provided in Fig. \ref{fig:All_D2O} where the time axis has been shifted to make time zero coincide with the valving in of the dryer. The curves were normalized to the average ending partial pressure values. It is initially clear that lower humidities flatten and extend the $D_2O$ breakthrough curve.

    \begin{wrapfigure}{r}{0.5\textwidth}
        \centering
        \includegraphics[width=0.5\columnwidth]{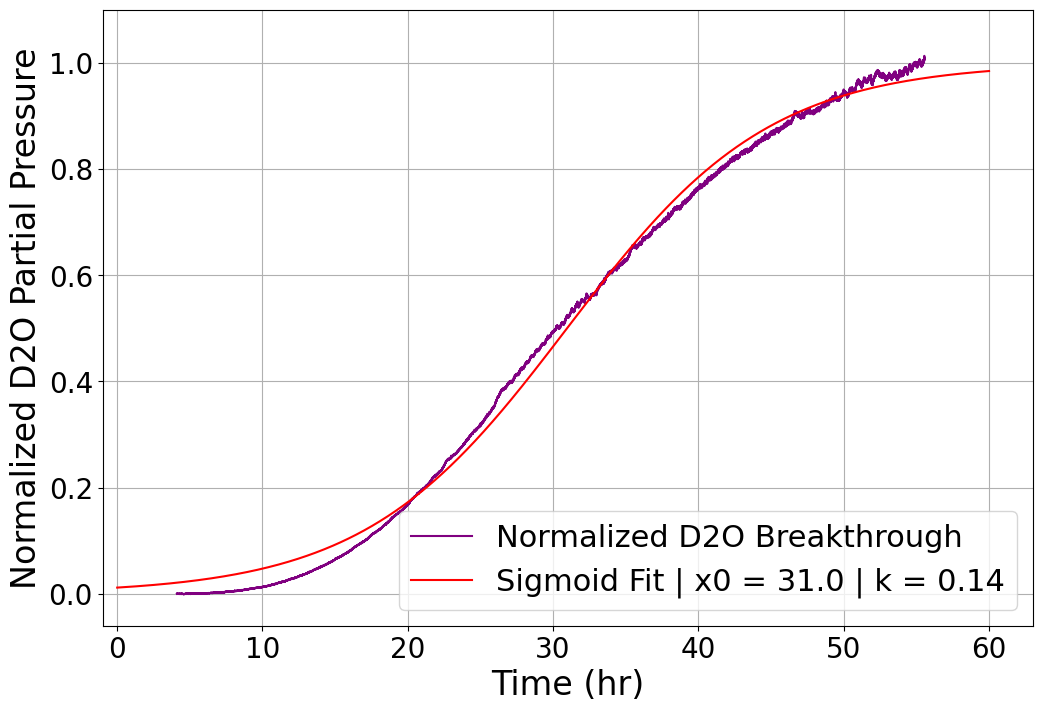}
        \caption{Sigmoid Fit for Trial 1 $D_2O$ partial pressure curve}
        \label{fig:SigFitExample}
    \end{wrapfigure}

    To better understand the evolution of the $D_2O$ partial pressure, each curve can be fit with a sigmoid curve (Eq. \ref{eq:Sigmoid}). Each $D_2O$ partial pressure curve has been normalized to span from 0 to 1 to  allow sigmoidal fitting with two fit parameters. Normalizing the trial 1 $D_2O$ partial pressure curve from Fig. \ref{fig:D2O_Trial1} and fitting a sigmoid is shown in Fig. \ref{fig:SigFitExample}.

\begin{figure}[h]
    \centering
    \begin{subfigure}[b]{0.45\textwidth} 
    \includegraphics[width=\textwidth]{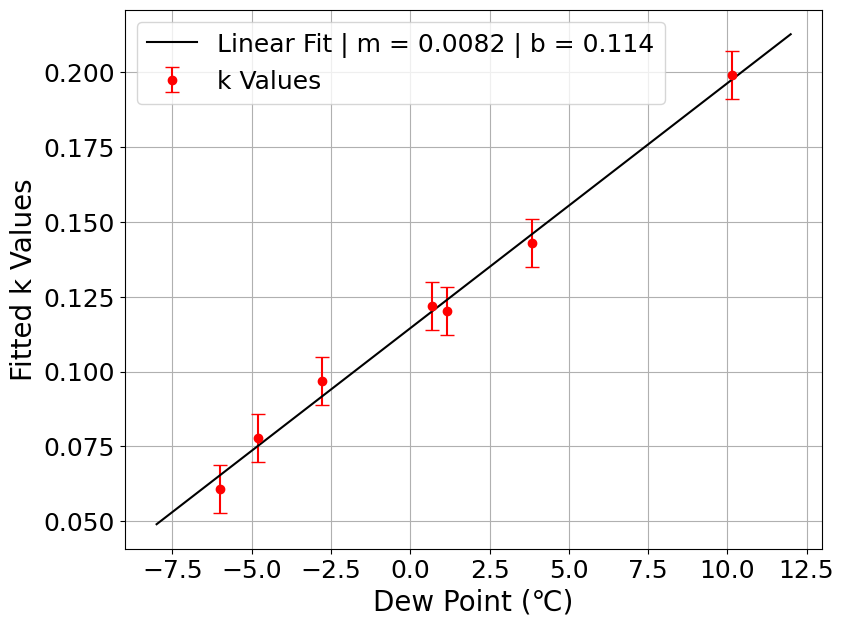}
    \caption{Fitted $k$ values against $D_2O$ dew point}
    \label{fig:k_values}
    \end{subfigure}
    \hspace{0.5 cm}
    \begin{subfigure}[b]{0.45\textwidth}
    \includegraphics[width=\textwidth]{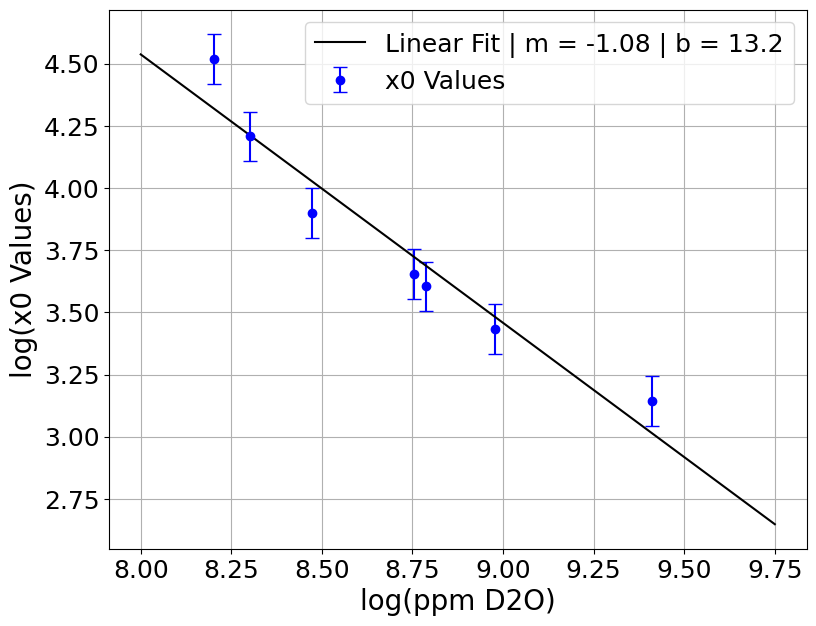}
    \caption{Logarithm of Fitted $x_0$ values and $D_2O$ ppm}
    \label{fig:x0_values}
    \end{subfigure}
    \caption{$D_2O$ Displacement Sigmoid Fit Parameters versus dew point/concentration of heavy water}
    \label{fig:SigmoidParameters}
\end{figure}

    Fitting all curves and plotting the $k$ values against the $D_2O$ dew point shows a linear relationship. This parameter represents the slope of the sigmoid curve. Higher concentrations of $D_2O$ flow increase the partial pressure rebound rate. These values are seen in Fig \ref{fig:k_values}. This simple linear relationship is not expected to hold for lower $D_2O$ concentrations because at extremely low concentrations the impact of the humidity must become vanishingly small. By setting $k=0$, as shown in Eq. \ref{eq:k_zero}. This current relationship predicts negative k values below -14°C, implying some non-linear behavior as concentrations decrease.

    \begin{equation}
        \centering
        k=0.0082\cdot D_P + 0.114 \Longrightarrow 0=0.0082\cdot D_{P_0} + 0.114 \Longrightarrow D_{P_0}=-13.9 ^\circ C
        \label{eq:k_zero}
    \end{equation}

    As an example, say there was a release of $500 Ci$ of tritium in a room of $5000 m^3$ volume. Dividing $500 Ci$ by the specific activity of tritium ($2.5886 \frac{Ci}{scc}$), we can find the volume of tritium within the room. A simple fraction will yield the concentration of $38.6ppb$, which, when we assume all tritium reacts to form tritium oxide, correlates to a dew point near -95°C. We can attempt to calculate the k value, as seen below, but this yields a negative and nonphysical value.

    \begin{equation}
        \centering
        k=0.0082\cdot \left(-95^\circ C\right) + 0.114 \Longrightarrow k=-0.665
        \label{eq:95DPEx}
    \end{equation}

    A similar procedure was performed with the $x_0$ fit parameter. In this case, the dew points are converted to ppm of $D_2O$, and then the logarithms are taken of both the ordinate and abscissa to yield a linear relationship. The $x_0$ parameter corresponds to the central point of the sigmoid curve, which, in this context, represents the number of hours after valve-in. A near-linear relationship was also recovered with this plotting procedure. This plot is shown in Fig. \ref{fig:x0_values}.

\section{Conclusion}
\label{sec:conclusion}
   Heavy water was shown to displace light water previously adsorbed into a molecular sieve 4a dryer bed. The eventual gradual increase in $D_2O$ concentration downstream of the dryer bed was adequately fit with sigmoid curves. This revealed useful linear relationships between the sigmoidal fit parameters and the humidity of the $D_2O$. 

    The application of dryers in fusion systems to collect tritiated water vapor will require operation at a potential concentration between 1 ppb and 1 ppm, much lower than the approximately 3600 ppm used in the lowest humidity $D_2O$ tests. 
    
\section*{Acknowledgments}

This experiment was funded in part by Torion Plasma Corporation and by Torion USA Inc. The authors acknowledge their contributions.

\pagebreak
\bibliographystyle{unsrtnat}
%custom ANS journal submission template bibliography style
\bibliography{bibliography}

\end{document}